\documentclass[
    ,final            
  ]
  {aipproc}

\layoutstyle{6x9}

\begin{document}

\title{Neutrino Mass and Flavour Models\footnote{Plenary talk at SUSY'09}}


\author{Stephen F King \footnote{e-mail: king@soton.ac.uk}}{
  address={School of Physics and Astronomy,
University of Southampton, Southampton SO17 1BJ, UK} }



\begin{abstract}
We survey some of the recent promising developments in the search
for the theory behind neutrino mass and tri-bimaximal mixing, and
indeed all fermion masses and mixing. We focus in particular on
models with discrete family symmetry and unification, and show how
such models can also solve the SUSY flavour and CP problems. We
also discuss the theoretical implications of the measurement of a
non-zero reactor angle, as hinted at by recent experimental
measurements.
\end{abstract}

\maketitle


\section{Introduction}
It has been one of the long standing goals of theories of particle
physics beyond the Standard Model (SM) to predict quark and lepton
masses and mixings. With the discovery of neutrino mass and
mixing, this quest has received a massive impetus. Indeed, perhaps
the greatest advance in particle physics over the past decade has
been the discovery of neutrino mass and mixing involving two large
mixing angles commonly known as the atmospheric angle
$\theta_{23}=43.1^o\pm 4^o$ and the solar angle
$\theta_{12}=34.5^o\pm 1.4^o$ where the current one sigma ranges
to typical global fits are displayed \cite{Schwetz:2008er}. There
is a $2\sigma $ hint for a non-zero reactor mixing angle
$\sin^2\theta_{13}=0.02\pm 0.01$ \cite{Fogli:2008jx} which gives
the one sigma range $\theta_{13}=8^o\pm 2^o$. The largeness of the
two large lepton mixing angles contrasts sharply with the
smallness of the quark mixing angles, and this observation,
together with the smallness of neutrino masses, provides new and
tantalizing clues in the search for the origin of quark and lepton
flavour which has led to a resurgence of interest in this subject \cite{Frampton:2004ud}.

It is a striking fact that current data on lepton mixing is
(approximately) consistent with the so-called tri-bimaximal (TB)
mixing pattern \cite{Harrison:2002er},
\begin{equation}
\label{TBM}
U_{TB}= \left(\begin{array}{ccc} \sqrt{\frac{2}{3}}& \frac{1}{\sqrt{3}}&0\\
-\frac{1}{\sqrt{6}}&\frac{1}{\sqrt{3}}&\frac{1}{\sqrt{2}}\\
\frac{1}{\sqrt{6}}&-\frac{1}{\sqrt{3}}&\frac{1}{\sqrt{2}}
\end{array} \right) P_{Maj},
\end{equation}
where $P_{Maj}$ is the diagonal phase matrix involving the two
observable Majorana phases. However there is no convincing reason
to expect exact TB mixing, and in general we expect deviations.
These deviations can be parametrized by three parameters $r,s,a$
defined as \cite{King:2007pr}:
\begin{equation}
\sin \theta_{13} = \frac{r}{\sqrt{2}}, \ \ \sin \theta_{12} =
\frac{1}{\sqrt{3}}(1+s), \ \ \sin \theta_{23} =
\frac{1}{\sqrt{2}}(1+a). \label{rsa}
\end{equation}
Global fits of the conventional mixing angles
\cite{Schwetz:2008er,Fogli:2008jx} can be translated into the
$1\sigma$ ranges
\begin{equation}
0.14<r<0.24,\ \ -0.05<s<0.02, \ \ -0.04<a<0.10.
\end{equation}
Note in particular that the central value of $r$ is now 0.2 which
corresponds to a 2$\sigma$ indication for a non-zero reactor angle
as discussed in \cite{Fogli:2008jx}.

Clearly a non-zero value of $r$, if confirmed, would rule out TB
mixing. However it is possible to preserve the good predictions
that $s=a=0$, by postulating a modified form of mixing matrix
called tri-bimaximal-reactor (TBR) mixing \cite{King:2009qt},
\begin{eqnarray}
U_{TBR} = \left( \begin{array}{ccc}
\sqrt{\frac{2}{3}}  & \frac{1}{\sqrt{3}} & \frac{1}{\sqrt{2}}re^{-i\delta } \\
-\frac{1}{\sqrt{6}}(1+ re^{i\delta })  & \frac{1}{\sqrt{3}}(1-
\frac{1}{2}re^{i\delta })
& \frac{1}{\sqrt{2}} \\
\frac{1}{\sqrt{6}}(1- re^{i\delta })  & -\frac{1}{\sqrt{3}}(1+
\frac{1}{2}re^{i\delta })
 & \frac{1}{\sqrt{2}}
\end{array}
\right)P_{Maj}. \label{MNS3}
\end{eqnarray}

\section{Neutrino Flavour Symmetry}
Let us expand the neutrino mass matrix in the diagonal charged
lepton basis, assuming exact TB mixing, as
${M^{\nu}_{TB}}=U_{TB}{\rm diag}(m_1, m_2, m_3)U_{TB}^T$ leading
to (absorbing the Majorana phases in $m_i$):
\begin{equation}
\label{mLL} {M^{\nu}_{TB}}= m_1\Phi_1 \Phi_1^T + m_2\Phi_2 \Phi_2^T + m_3\Phi_3 \Phi_3^T
\end{equation}
where $\Phi_1^T=\frac{1}{\sqrt{6}}(2,-1,1)$,
$\Phi_2^T=\frac{1}{\sqrt{3}}(1,1,-1)$, $\Phi_3^T=\frac{1}{\sqrt{2}}(0,1,1)$,
are the respective columns of $U_{TB}$
and $m_i$ are the physical neutrino masses. In the neutrino
flavour basis (i.e. diagonal charged lepton mass basis), it has
been shown that the above TB neutrino mass matrix is invariant
under $S,U$ transformations:
\begin{equation}
{M^{\nu}_{TB}}\,= S {M^{\nu}_{TB}} S^T\,= U {M^{\nu}_{TB}} U^T \ .
\label{S} \end{equation}
A very straightforward argument
\cite{King:2009mk} (see also \cite{Lam:2008sh,Grimus:2009pg})
shows that this neutrino flavour symmetry group
has only four elements corresponding to Klein's four-group $Z_2^S
\times Z_2^U$. By contrast the diagonal charged lepton mass matrix
(in this basis) satisfies a diagonal phase symmetry $T$. The
matrices $S,T,U$ form the generators of the group $S_4$ in the
triplet representation, while the $A_4$ subgroup is generated by
$S,T$.

\section{Family Symmetry: Direct vs Indirect Models}
As discussed in \cite{King:2009ap},
the flavour symmetry of the neutrino mass matrix may originate
from two quite distinct classes of models. The first class of models,
which we call direct models, are based on a family symmetry $G_f=S_4$,
or a closely related family symmetry as discussed below,
some of whose generators are directly preserved in the
lepton sector and are manifested as part of the observed flavour
symmetry. The second class of models, which we call indirect
models, are based on some more general family symmetry $G_f$ which is completely
broken in the neutrino sector, while the observed neutrino flavour
symmetry $Z_2^S \times Z_2^U$ in the neutrino flavour basis emerges as an
accidental symmetry which
is an indirect effect of the family symmetry $G_f$. In such
indirect models the flavons responsible for the neutrino masses break $G_f$
completely so that none of the generators of $G_f$ survive in the
observed flavour symmetry $Z_2^S \times Z_2^U$.

\begin{figure}[h]
\includegraphics[width=24pc]{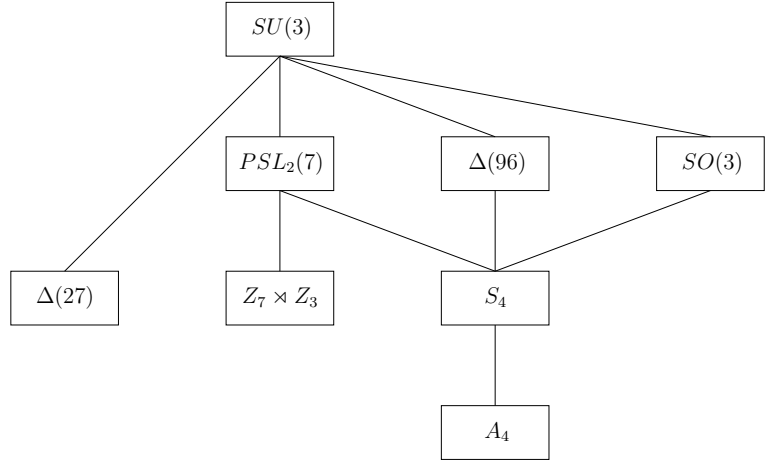}
\caption{\label{family}Some possible family symmetry groups.}
\end{figure}

In the direct models, the symmetry of the neutrino mass matrix in
the neutrino flavour basis (henceforth called the neutrino mass matrix for
brevity) is a remnant of the $G_f=S_4$ symmetry of the Lagrangian,
where the generators
$S,U$ are preserved in the neutrino sector, while the diagonal generator
$T$ is preserved in the charged lepton sector.
For direct models, a larger family symmetry $G_f$ which
contains $S_4$ as a subgroup is also possible e.g. $G_f=PSL(2,7)$ \cite{King:2009mk}.
Some possible family symmetry groups and their relation to $S_4$ are shown in
Figure~\ref{family}.
If the family symmetry of the underlying
Lagrangian is smaller, say, $G_f=A_4$ \cite{Ma:2001dn}, then in some cases this can lead to a direct
model where the $T$ generator of the underlying Lagrangian
symmetry is preserved in the charged lepton sector, while the $S$
generator is preserved in the neutrino sector, with the $U$
transformation of $S_4$ emerging as an accidental symmetry due to the
absence of flavons in the ${\bf 1',1''}$ representations of $A_4$
\cite{Altarelli:2005yp}. Typically direct models satisfy form
dominance \cite{Chen:2009um}, and require flavon F-term
vacuum alignment, permitting an $SU(5)$ type unification
\cite{Altarelli:2005yp}. Such minimal $A_4$ models lead to
neutrino mass sum rules between the three masses $m_i$, resulting
in/from a simplified mass matrix in Eq.\ref{mLL}. $A_4$ may result
from 6D orbifold models \cite{Altarelli:2006kg} and recently a 6D
$A_4\times SU(5)$ SUSY GUT model has been constructed \cite{Burrows:2009pi}.

In the indirect models \cite{King:2009ap} the idea is that the three columns of $U_{TB}$
$\Phi_i$ are promoted to new Higgs fields called
``flavons'' whose VEVs break the family symmetry, with the
particular vacuum alignments along the directions
$\Phi_i$. In the indirect models
the underlying family symmetry of the Lagrangian $G_f$ is
completely broken, and the flavour symmetry of the neutrino mass
matrix $Z_2^S \times Z_2^U$ emerges entirely as an accidental
symmetry, due to the presence of flavons with particular vacuum
alignments proportional to the columns of $U_{TB}$, where such
flavons only appear quadratically in effective Majorana
Lagrangian \cite{King:2009ap}. Such vacuum alignments can be
elegantly achieved using D-term vacuum alignment, which allows
the large classes of discrete family symmetry $G_f$,
namely the $\Delta(3n^2)$ and $\Delta(6n^2)$ groups \cite{King:2009ap}.

\section{See-saw mechanism and form dominance}
It is possible to derive the TB form of the neutrino mass matrix
in Eq.\ref{mLL} from the see-saw mechanism in a very elegant way
as follows. In the diagonal right-handed neutrino mass basis we
may write $M_{RR}^{\nu}={\rm diag}(M_A, M_B, M_C)$ and the Dirac
mass matrix as $M_{LR}^{\nu}=(A,B,C)$ where $A,B,C$ are three
column vectors. Then the type I see-saw formula
${M^{\nu}}=M_{LR}^{\nu}(M_{RR}^{\nu})^{-1}(M_{LR}^{\nu})^T$
gives
\begin{equation}
\label{mLLCSD} {M^{\nu}}=\frac{AA^T}{M_A}+ \frac{BB^T}{M_B} +
\frac{CC^T}{M_C}.
\end{equation}
By comparing Eq.\ref{mLLCSD} to the TB form in Eq.\ref{mLL} it is
clear that TB mixing will be achieved if $A\propto \Phi_3$,
$B\propto \Phi_2$, $C\propto \Phi_1$, with each of $m_{3,2,1}$
originating from a particular right-handed neutrino of mass
$M_{A,B,C}$, respectively. This mechanism
allows a completely general neutrino mass spectrum and, since the
resulting ${M^{\nu}}$ is form diagonalizable, it is referred to
as form dominance (FD) \cite{Chen:2009um}. For example, it has
recently been show that the direct $A_4$ see-saw models
\cite{Altarelli:2005yp} satisfy FD \cite{Chen:2009um}, where each
column corresponds to a linear combination of flavon VEVs.

A more natural possibility, called Natural FD, arises when each column
arises from a separate flavon VEV, and this possibility corresponds to
the case of indirect models. For example,
if $m_1\ll m_2 < m_3$ then the precise form of $C$ becomes
irrelevant, and in this case FD reduces to constrained sequential
dominance (CSD)\cite{King:2005bj}. The CSD mechanism has been
applied in this case to the class of indirect models
with Natural FD based on the family symmetries
$SO(3)$ \cite{King:2005bj,King:2006me} and $SU(3)$
\cite{deMedeirosVarzielas:2005ax}, and their discrete subgroups
\cite{deMedeirosVarzielas:2005qg}.

\section{Partially Constrained sequential dominance}

It is possible to achieve TBR mixing, corresponding to $s=a=0$ but
$r\neq 0$, by a slight modification to the CSD conditions,
\begin{equation}
\label{PCSD} B = \frac{b}{\sqrt{3}} \left(
\begin{array}{r}
1 \\
1 \\
-1
\end{array}
\right),\ \ A  = \frac{c}{\sqrt{2}} \left(
\begin{array}{r}
\varepsilon \\
1 \\
1
\end{array}
\right).
\end{equation}
We refer to this as Partially Constrained Sequential Dominance
(PCSD)\cite{King:2009qt}, since one of the conditions of CSD is
maintained, while the other one is violated by the parameter
$\varepsilon$. Note that the introduction of the parameter
$\varepsilon$ also implies a violation of FD since the columns of
the Dirac mass matrix $A,B$ can no longer be identified with the
columns of the MNS matrix, due to the non-orthogonality of $A$ and
$B$. To leading order in $|m_2|/|m_3|$ the mass matrix resulting
from Eq.\ref{PCSD} leads to TBR mixing where we identify
\cite{King:2009qt},
\begin{equation}
m_1=0, \ \ m_2=b^2/M_B, \ \ m_3=a^2/M_A, \ \ \varepsilon =
re^{-i\delta }.
\end{equation}
Thus, the TBR form of mixing matrix in Eq.\ref{MNS3} will result,
to leading order in $|m_2|/|m_3|$.
For example, it is straightforward to implement the above example
of PCSD into realistic GUT models with non-Abelian family symmetry
spontaneously broken by flavons which are based on the CSD
mechanism
\cite{King:2005bj,deMedeirosVarzielas:2005ax,King:2006me}.
More generally it is natural to expect vacuum alignments as in Eq.\ref{PCSD}
from the D-term vacuum alignment associated with the indirect models \cite{King:2009ap}.
The point is that D-term alignment along the direction of the second column of $U_{TB}$ $\Phi_2$
is enforced by the family symmetry $G_f$, but the other alignments are provided by orthogonality
arguments and are therefore intrinsically more model dependent.

\section{Family Symmetry $\otimes$ GUT models}

In typical Family Symmetry $\otimes$ GUT models the origin of the
quark mixing angles derives predominantly from the down quark
sector, which in turn is closely related to the charged lepton
sector. There are many possibilities for the choice of family symmetry and
GUT symmetry. Examples of indirect models along these lines include the Pati-Salam gauge group
$SU(4)_{PS}\times SU(2)_L\times SU(2)_R$ in combination with
$SU(3)$ \cite{deMedeirosVarzielas:2005ax}, $SO(3)$
\cite{King:2005bj,King:2006me}, $A_4$ \cite{King:2006np} or
$\Delta_{27}$ \cite{deMedeirosVarzielas:2006fc}. Examples of direct models
along these lines are based on $SU(5)$ GUTs in combination with $A_4$
\cite{Altarelli:2008bg} or $T'$ \cite{Chen:2007afa}.

\begin{figure}[h]
\includegraphics[width=24pc]{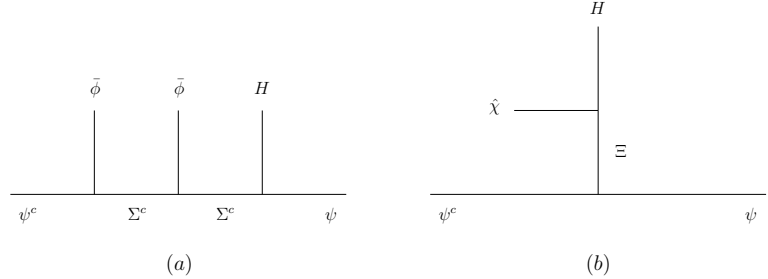}
\caption{\label{diagrams}Diagrams responsible for the Yukawa couplings in the
$PSL(2,7)\times SO(10)$ model. Diagram (a) shows how two triplet flavon $\phi$ insertions are responsible
for the first and second family Yukawa couplings, while diagram (b) shows how a single sextet flavon
$\chi$ insertion is responsible for the third family Yukawa couplings.}
\end{figure}

A promising new example of a Family Symmetry $\otimes$ GUT model is the
$PSL(2,7)\times SO(10)$ proposal in \cite{King:2009mk}.
Such a model unifies the three families into a complex $\psi \sim (3,16)$
representation, with the Higgs $H \sim (1,10)$,
while obtaining the third family Yukawa coupling
from a sextet flavon $\chi \sim (6,1)$ coupling $\chi \psi \psi H$.
The other Yukawa couplings arise from two triplet flavons $\phi$.
The diagrams responsible for the Yukawa couplings are shown in Fig.\ref{diagrams}.

In order to reconcile the down quark and charged lepton
masses, simple ansatze, such as the Georgi-Jarlskog hypothesis
\cite{Georgi:1979df}, lead to very simple approximate expectations
for the charged lepton mixing angles such as $\theta^e_{12}\approx
\lambda/3$, $\theta^e_{23}\approx \lambda^2$,
$\theta^e_{13}\approx \lambda^3$, where $\lambda \approx 0.22$ is
the Wolfenstein parameter from the quark mixing matrix. If the
family symmetry enforces accurate TB mixing in the neutrino
sector, then $\theta^e_{12}\approx \lambda/3$ charged lepton
corrections will cause deviations from TB mixing in the physical
lepton mixing angles, and lead to a sum rule relation
\cite{King:2005bj,Masina:2005hf,Antusch:2005kw}, which can be
conveniently expressed as \cite{King:2007pr} $s\approx r \cos
\delta$ where $r\approx \lambda /3$ and $\delta$ is the observable
CP violating oscillation phase, with RG corrections of less than
one degree \cite{Boudjemaa:2008jf}. Such sum rules can be tested
in future high precision neutrino oscillation experiments
\cite{Antusch:2007rk}.

Note that in such a GUT-flavour framework, one expects the charged
lepton corrections to the neutrino mixing angles to be less than
of order $\theta_{12}^e/\sqrt{2}$ (where typically $\theta_{12}^e$
is a third of the Cabibbo angle) plus perhaps a further $1^o$ from
renormalization group (RG) corrections. Thus such theoretical
corrections cannot account for an observed reactor angle as large
as $8^o$, corresponding to $r=0.2$, starting from the hypothesis
of exact TB neutrino mixing.


\section{Family symmetry and SUSY Flavour/CP problems}

In SUSY models we not only want to understand the origin of the Yukawa couplings,
but also the soft SUSY breaking masses. There are stringent limits from
of flavour changing and CP violating processes on the form of these soft masses.
These limits may be expressed as bounds on the real and imaginary parts of
the mass insertion parameters $\delta$, as recently compiled in \cite{Antusch:2007re}.
It has been observed that $G_f=SU(3)$ family symmetry implies an approximately universal family
structure of the soft masses close to the GUT scale \cite{Ross:2004qn} and, with the hypothesis that
CP is spontaneously broken by flavon VEVs, this can also lead to suppressed CP violation \cite{Ross:2004qn}.
These issues have been recently examined in detail in the $G_f=SU(3)$ models which predict tri-bimaximal mixing
\cite{Antusch:2007re}, although the results are also applicable to discrete family symmetry models
such as the $G_f=\Delta (27)$ model  \cite{deMedeirosVarzielas:2006fc}. The results show that there is a small
tension in the model (at least for ``reasonable'' SUSY masses and parameters)
due to the processes $\mu \rightarrow e \gamma$ and the EDMs \cite{Antusch:2007re}.
However this tension can be completely removed in classes of $G_f=SU(3)$ models based on supergravity \cite{Antusch:2008jf}.

\section{Conclusion}
We have surveyed some of the recent promising developments in the search
for the theory behind neutrino mass and tri-bimaximal mixing, and
indeed all fermion masses and mixing. Tri-bimaximal mixing implies
a discrete neutrino flavour symmetry $Z_2^S \times Z_2^U$ which can be realized either directly or indirectly
via a discrete family symmetry $G_f$. The direct models are typically based on $G_f=A_4$ or $G_f=S_4$
where the family symmetry generators $S$ and $U$ are preserved in the neutrino sector.
The indirect models are typically based on $G_f=\Delta(3n^2)$ or $G_f=\Delta(6n^2)$
where none of the family symmetry generators are preserved in the neutrino sector since they are all broken
by the flavons which align along the columns of $U_{TB}$. However (after the see-saw mechanism) the neutrino sector
involves only quadratic combinations of these flavons leading to an accidental
discrete neutrino flavour symmetry $Z_2^S \times Z_2^U$.

The type I see-saw mechanism can be elegantly implemented using form dominance in which the
columns of the neutrino Yukawa matrix (in the diagonal charged lepton and right-handed neutrino mass
basis) are proportional to the columns of $U_{TB}$. In the direct models these columns are generated
from linear combinations of different flavon alignments, which implies a mild tuning
of VEVs to achieve an acceptable neutrino mass pattern. In the indirect models, each column is
generated from a unique flavon aligned along a direction corresponding to a column of $U_{TB}$ so there
is no tuning required for the neutrino masses, and this is called natural form dominance.
In the limit that $m_1\ll m_2 < m_3$ natural form dominance reduces to constrained sequential
dominance.

We have also discussed the theoretical implications of the measurement of a
non-zero reactor angle, as hinted at by recent experimental
measurements. A measurement of a
large reactor angle, consistent with the present $2\sigma$
indication for $r=0.2$, can still be consistent with tri-bimaximal
solar and atmospheric mixing, corresponding to $s=a=0$, according
to the tri-bimaximal-reactor mixing hypothesis. This can be achieved in the see-saw mechanism
using partially constrained sequential dominance, which may be readily be realized in classes
of indirect models based on D-term vacuum alignment.

We have surveyed models based on a Family Symmetry $\otimes$ GUT symmetry structure
and noted that direct models tend to be based on $SU(5)$ GUTs while indirect models
allow $SO(10)$ GUTs. Another more technical distinction is that direct models are based on
F-term vacuum alignment, while indirect models often use D-term vacuum alignment.
However the new class of models based on $PSL(2,7)$ \cite{King:2009mk}
involving sextet flavons (allowing improved top quark Yukawa convergence)
may yield a direct model involving D-term vacuum alignment
and $SO(10)$.

Finally we have noted that models based on non-Abelian family symmetry
can solve the SUSY flavour and CP problems. This increases the motivation
for considering models with non-Abelian family symmetry, especially
models in which all three entire quark and lepton families (including both left and
right-handed components) transform
as a triplet under the family symmetry, since such models offer maximum
protection against SUSY induced flavour changing and CP violation.




\begin{theacknowledgments}
My attendance at this conference was made possible by a grant from the Royal Society, London, UK.
I would also like to thank the organizers for their hospitality.
\end{theacknowledgments}







\end{document}